\documentstyle[titlepage,12pt]{article}

\begin{document}
\title {Propagation of perturbations along strings}
\vspace{5 in}
\author{A.L.Larsen\\NORDITA, Blegdamsvej 17, DK-2100 Copenhagen \O,
Denmark\thanks{E-mail: allarsen@nbivax.nbi.dk}\and V.P.Frolov\thanks{On
leave of absence from P.N.Lebedev Physical Institute, Leninsky prospect 53,
\hspace*{7mm}Moscow 117924 Russia}\\The Copenhagen University Observatory,\\
\O ster Voldgade 3, DK-1350 Copenhagen K, Denmark\thanks{E-mail:
frolov@astro.ku.dk}}
\maketitle
\begin{abstract}
A covariant formalism for physical perturbations propagating along a string
in an
arbitrary curved spacetime is developed. In the case of a stationary string
in a static background the propagation of the perturbations is described
by a
wave-equation with a potential consisting of 2 terms: The first term describing
the time-dilation and the second is connected with the curvature
of space. As applications of the developed approach the propagation of
perturbations along a stationary string in Rindler, de Sitter, Schwarzschild
and
Reissner-Nordstr\"{o}m spacetimes are investigated.
\end{abstract}
\newpage
\section{Introduction}
Recently there has been a lot of interest in string theory formulated in curved
spacetime. The main complication, as compared to the case of flat Minkowski
space, is related to the non-linearity of the equations of motion. It makes it
possible to obtain a complete analytic solution only in a very few cases.
These special cases include maximally symmetric spacetimes [1],
conical spacetime [2] and
gravitational shock-wave background [3]. Besides these only some special
explicit analytic solutions are known. The consideration of small perturbations
on their background is of interest, e.g. it allows one to
investigate the stability-properties of these solutions [4].
More generally it is interesting to see how small waves propagate
on a string or a membrane [5], and especially to see how they
are affected by the curvature of space and time.

Small perturbations may play an important role for cosmic strings [6].
Cosmic strings may gain a
significant part of their energy from their small scale structure. The effect
of
such small scale "wiggles" on the equation of state of a cosmic string has
been investigated by various authors [7], and also more mathematical aspects
concerning separability of the equations of motion  of "wiggly" strings
and "noisy" strings in curved spacetimes [8] has been studied.

Finally string perturbations can be invoked for the purpose of a
"semiclassical"
quantization procedure for strings in curved spacetime, by taking a classical
solution to the equations of motion and then considering small perturbations
as quantum fluctuations. This approach was developed by Vega and S\'{a}nches
in Ref. 9, with application to various gravitational backgrounds
including black holes, de Sitter space, Rindler space and several
others.

The main aim of the present paper is to develop the general theory of small
perturbations
propagating along a string in curved spacetime.

In section 2 we consider the first and second variations of the Polyakov
action [10]. The first variation of the action gives the equations of motion
for the string in a curved background,
while the second variation can be used to obtain the equations describing
the perturbations propagating in the background
of the exact solution. From a geometrical point of view the problem we are
interested in is the study of
the variations of a minimal 2-surface embedded in a curved 4-space.
In section 3 we consider perturbations propagating along a stationary string in
a static
spacetime. We exploit the fact that such a stationary configuration of the
string
can be described by a geodesic equation in a properly chosen 3-dimensional
("unphysical") space [11]. This
approach allows one to simplify general results of section 2, and to
obtain a relatively simple "wave-equation" determining the propagation of
perturbations.
In section 4 we give a few examples to illustrate our general results. We
consider perturbations propagating along stationary strings in Rindler,
de Sitter, Schwarzschild and Reissner-Nordstr\"{o}m spacetimes, as well as in
the
quasi-Newtonian gravitational field. As a
curiosity we find that the perturbations propagating along a
curved string in (flat) Rindler space and the perturbations
propagating along a straight string in (curved) de Sitter space
is determined by one and the same well-known equation from quantum mechanics:
The P\"{o}schl-Teller equation.

In the paper we use sign-conventions of  Misner,Thorne,Wheeler [12].
\section{General physical perturbations}
In this section we derive an effective action and
equations of motion for small perturbations on an arbitrary
string-configuration
in an arbitrary 4-dimensional gravitational background.

Our starting point is the Polyakov action for the relativistic string [10]:
\begin{equation}
{\cal S}=\int d\tau d\sigma\sqrt{-h}h^{AB}G_{AB}.
\end{equation}
We use units in which beside $G=1$, $c=1$ the string tension
$(2\pi\alpha')^{-1}=1$.
In Eq. (2.1) $h_{AB}$ is the internal metric with determinant $h$
whereas $G_{AB}$ is the induced metric
on the world-sheet:
\begin{equation}
G_{AB}=g_{\mu\nu}\frac{\partial x^\mu}{\partial\xi^A}\frac{\partial x^\nu}{
\partial\xi^B}=g_{\mu\nu}x^\mu_{,A}x^\nu_{,B}.
\end{equation}
$x^\mu$ ($\mu=0,1,2,3$) are the spacetime coordinates and
$\xi^A$ ($A=0,1$) are the world-sheet coordinates
$(\xi^0,\xi^1)=(\tau,\sigma)$.
Finally $g_{\mu\nu}$ is the spacetime metric which in this section is
completely
arbitrary.

We now make variations $\delta x^\mu$ and $\delta h_{AB}$. The corresponding
variation of the action (2.1) is conveniently written in the form:
\begin{equation}
\delta{\cal S}=\int d\tau d\sigma\sqrt{-h}
\left( [\frac{1}{2}h^{AB}G^C\hspace{.5mm}_C-G^{AB}]
\delta h_{AB}-2g_{\mu\nu}[\Box x^\nu+h^{AB}\Gamma^\nu_{\rho\sigma}x^\rho_{,A}
x^\sigma_{,B}]\delta x^\mu\right),
\end{equation}
where $G^C\hspace{.5mm}_C=h^{BC}G_{BC}$ is the trace
of the induced metric on the world-sheet and $\Box$ is the d'Alambertian:
\begin{equation}
\Box=\frac{1}{\sqrt{-h}}\partial_A (\sqrt{-h}h^{AB}\partial_B).
\end{equation}
{}From $\delta{\cal S}$ we get the usual equations of motion:
\begin{equation}
G_{AB}-\frac{1}{2}h_{AB}G^C\hspace{.5mm}_C=0,
\end{equation}
\begin{equation}
\Box x^\mu+h^{AB}\Gamma^\mu_{\rho\sigma}x^\rho_{,A}x^\sigma_{,B}=0.
\end{equation}
Equation (2.5) shows that the 2-dimensional world-sheet
energy-momentum tensor vanishes. This equation is to be considered
as a constraint on the solution
of the equation of motion for $x^\mu$ (2.6).

The equations describing the propagation of perturbations in the background
of an exact solution to Eqs. (2.5)-(2.6) can be obtained by variation of
these equations. Technically it
appears to be easier to make
the second variation of ${\cal S}$ by varying Eq. (2.3).
We use this approach. Let $x^\mu=x^\mu (\xi^A)$ be a solution to Eqs. (2.5) and
(2.6) describing a background strings configuration. In what follows it is
convenient to
introduce 2 vectors $n^\mu_R (R=2,3)$ normal to the surface of the
string world-sheet:
\begin{equation}
g_{\mu\nu}n^\mu_{R}n^\nu_S=\delta_{RS},\hspace{10mm}g_{\mu\nu}x^\mu_{,A}
n^\nu_R=0.
\end{equation}
The general perturbation $\delta x^\mu$ can be decomposed as:
\begin{equation}
\delta x^\mu=\delta x^R n^\mu_R+\delta x^A x^\mu_{,A}.
\end{equation}
It can be verified that the variations $\delta x^A x^\mu_{,A}$ leave ${\cal S}$
unchanced. This is the result of the invariance of the string action with
respect to reparametrizations of the world-sheet.
That is why we restrict ourselves by preserving in the variation of the
action only physical perturbations
$\delta x^\mu$, which can be written as:
\begin{equation}
\delta x^\mu=\delta x^R n^\mu_R.
\end{equation}
Note that $R=(2,3)$ is just a label for the normalvectors and summation over
repeated indices is implied.
Before carrying out the corresponding variation of Eq. (2.3) it is convenient
to
introduce the second fundamental form $\Omega_{R,AB}$ and the normal
fundamental
form $\mu_{RS,A}$ [13] defined for a given configuration of the strings
world-sheet:
\begin{equation}
\Omega_{R,AB}=g_{\mu\nu}n^\mu_R x^\rho_{,A}\nabla_\rho x^\nu_{,B},
\end{equation}
\begin{equation}
\mu_{RS,A}=g_{\mu\nu}n^\mu_R x^\rho_{,A}\nabla_\rho n^\nu_S,
\end{equation}
where $\nabla_\rho$ is the spacetime covariant derivative.
Note that $\Omega_{R,AB}=\Omega_{R,BA}$ whereas $\mu_{RS,A}=-\mu_{SR,A}$. Now
the
equations of motion (2.6) can be rewritten compactly as:
\begin{equation}
h^{AB}\Omega_{R,AB}=0,
\end{equation}
which is the well-known result [13] that a minimal surface has vanishing normal
curvature in all directions normal to the strings world-sheet. One can also
verify the following useful relation:
\begin{equation}
\delta G_{AB}=-2\Omega_{R,AB}\delta x^R.
\end{equation}

The variation of Eq. (2.3) turns out to be somewhat complicated leading to
quadratic
terms in $\delta x^R$ and $\delta h_{AB}$ as well as to mixed terms. After some
algebra one finds:
\begin{eqnarray}
\delta^2{\cal S}=\int d\tau d\sigma\sqrt{-h}\hspace*{-2mm}& ( &
\hspace*{-2mm} \delta h_{AB}
[2G^{BC}h^{AD}-\frac{1}{2}h^{AD}h^{BC}G^E\hspace*{.5mm}_E-
\frac{1}{2}h^{AB}G^{CD}]
\delta h_{CD}
\nonumber \\& + &\hspace*{-2mm}4\delta h_{AB}h^{AC}h^{BD}\Omega_{R,CD}
\delta x^R\nonumber \\
& - &\hspace*{-2mm}2\delta x^R [\delta_{RS}\Box-h^{AB}g_{\mu\nu}
(x^\rho_{,A}\nabla_\rho n^\mu_R)
(x^\sigma_{,B}\nabla_\sigma n^\nu_S)\nonumber \\& - &\hspace*{-2mm}2h^{AB}
\mu_{RS,A}\partial_B-
h^{AB}x^\mu_{,A}x^\nu_{,B}R_{\mu\rho\sigma\nu}n^\rho_R n^\sigma_S]
\delta x^S\hspace*{1mm}),
\end{eqnarray}
where $R_{\mu\rho\sigma\nu}$ is the Riemann curvature tensor in the spacetime
in which the string is embedded. The variation
of the internal metric
$\delta h_{AB}$ obeys the equation:
\begin{equation}
\Omega_{R,AB}\delta x^R=-\frac{1}{4}(G^C\hspace*{.5mm}_C\delta
h_{AB}-h_{AB}G^{CD}\delta h_{CD}).
\end{equation}
This relation gives:
\begin{equation}
\delta
h_{AB}\Omega_R\hspace*{.5mm}^{AB}=-\frac{4}{G^C\hspace*{.5mm}_C}\Omega_{R,AB}
\Omega_S\hspace*{.5mm}^{AB}\delta x^S.
\end{equation}
Eq. (2.16) show that $\delta h_{AB}$ is not a dynamical field and it can be
expressed in the algebraic way in terms of the perturbations $\delta x^S$. By
using Eq. (2.16) we can exclude the perturbations $\delta h_{AB}$ from the
variation of the action. By using Eq. (2.16) and the identity:
\begin{equation}
h^{AB}g_{\mu\nu}(x^\rho_{,A}\nabla_\rho n^\mu_R)(x^\sigma_{,B}\nabla_\sigma
n^\nu_S)=\mu_{RT}\hspace*{1mm}^A\mu_S\hspace*{1mm}^T\hspace*{1mm}_A+\frac{2}
{G^C\hspace*{.5mm}_C}\Omega_R\hspace*{.5mm}^{AB}\Omega_{S,AB},
\end{equation}
we finally obtain the effective action for the physical perturbations in the
form:
\begin{eqnarray}
{\cal S}_{eff}=\int d\tau d\sigma\sqrt{-h}\delta x^R(\delta_{RS}\Box
\hspace*{-2mm}&+&\hspace*{-2mm}2\mu_{RS}\hspace*{1mm}^A\partial_A-
\mu_{RT}\hspace*{1mm}^A\mu_S\hspace*{1mm}^T\hspace*{1mm}_A+
\frac{2}{G^C\hspace*{.5mm}_C}\Omega_R\hspace*{.5mm}^{AB}\Omega_{S,AB}\nonumber\\
\hspace*{-2mm}&-&\hspace*{-2mm}h^{AB}x^\mu_{,A}x^\nu_{,B}R_{\mu\rho\sigma\nu}
n^\rho_R n^\sigma_S) \delta x^S.
\end{eqnarray}
This is the main result of this section. Note that each of the terms of
equation (2.18) is spacetime
and world-sheet invariant independently. It should be stressed that the
various terms are not completely independent. It is well-known in differential
geometry that there exists relations between the second fundamental form, the
normal fundamental form and the Riemann tensor: The so-called Gauss-equation,
the Codazzi-Mainardi-equation and the Ricci-equation [13]. These relations,
however, do not seem to simplify the action (2.18).
The equations of motion corresponding to the action (2.18) are:
\begin{eqnarray}
\Box\delta x_R\hspace*{-2mm}&+&\hspace*{-2mm}2\mu_{RS}\hspace*{1mm}^A
(\delta x^S)_{,A}+(\nabla_A\mu_{RS}\hspace*{1mm}^A)\delta x^S-
\mu_{RT}\hspace*{1mm}^A\mu_S\hspace*{1mm}^T\hspace*{1mm}_A\delta x^S\nonumber
\\
\hspace*{-2mm}&+&\hspace*{-2mm}\frac{2}{G^C\hspace*{.5mm}_C}
\Omega_R\hspace*{.5mm}^{AB}\Omega_{S,AB}\delta x^S-h^{AB}x^\mu_{,A}x^\nu_{,B}
R_{\mu\rho\sigma\nu}n^\rho_R n^\sigma_S\delta x^S=0,
\end{eqnarray}
where $\nabla_A$ is the strings world-sheet covariant derivative. In general
this is a
complicated set of coupled partial (linear) second order differential
equations. In the following sections we will however show that the solution can
be
found analytically in various interesting cases.

We conclude this section
with the following remark. Obviously
there is an ambiguity in the choice of normalvectors $n^\mu_R$ introduced in
Eq. (2.7) connected with
their local "rotations":
\begin{equation}
n^\mu_R\longrightarrow n^\mu_R+\lambda (\tau,\sigma)\epsilon_R\hspace*{1mm}^S
n_S^\mu,
\end{equation}
\begin{equation}
\delta x^R\longrightarrow\delta x^R+\lambda (\tau,\sigma)\epsilon^R
\hspace*{1mm}_S\delta x^S,
\end{equation}
where $\lambda (\tau,\sigma)$ is an arbitrary (small) function and
$\epsilon_{RS}$ is the antisymmetric Levi-Civita symbol. The effective action
(2.18) remains invariant under these local rotations. Note that the last 2
terms in the bracket of Eq. (2.18) are invariants separately whereas the first
3 terms
must be taken together. The invariance with respect to transformations (2.20)
and (2.21) can be used to simplify the
action by a suitable choice of the normalvectors. In the next section we will
show that for a stationary string in a static background one can get rid of
the terms in Eq. (2.18) involving the normal fundamental form $\mu_{RS,A}$ by
using this gauge freedom.
\vskip 6pt
\section{Stationary string in static background}
\setcounter{equation}{0}
We will now specialize to stationary strings in static spacetimes, i.e. we
take a stationary solution to Eqs. (2.5)-(2.6) and consider time-dependent
perturbations around it.

The metric of a static spacetime can in general
be written:
\begin{equation}
g_{\mu\nu}=\left( \begin{array}{cc} -F & 0 \\ 0 & H_{ij}/F \end{array}
\right),
\end{equation}
where $\partial_t F=0$, $\partial_t H_{ij}=0$ and $i,j=1,2,3$. The
nonvanishing components of the Christoffel symbol for the metric $g_{\mu\nu}$
are $\Gamma^i_{jk}$ and:
\begin{equation}
\Gamma^0_{i0}=\frac{F_{,i}}{2F},\hspace*{5mm}\Gamma^i_{00}=
\frac{F}{2}H^{ij}F_{,j}.
\end{equation}
The solution describing a stationary string can be parametrized
in the following way:
\begin{equation}
t=x^0=\tau,\hspace*{5mm}x^i=x^i(\sigma),
\end{equation}
so that:
\begin{equation}
x^\mu_{,0}=(1,0,0,0),\hspace*{5mm}x^\mu_{,1}=(0,x'^i).
\end{equation}
(The prime denotes the differentiation with respect to $\sigma$.)
For this parametrization the induced metric on the world-sheet is:
\begin{equation}
G_{00}=-F,\hspace*{5mm}G_{01}=G_{10}=0,\hspace*{5mm}G_{11}=\frac{H_{ij}}{F}
x'^i x'^j.
\end{equation}
Working
with the Nambu-Goto action [14] we identify
the internal metric and the induced metric:
\begin{equation}
G_{AB}=h_{AB}.
\end {equation}
The equation of motion (2.6) in this case reduces to:
\begin{equation}
x''^i+\tilde{\Gamma}^i_{jk}x'^j x'^k=0,
\end{equation}
where $\tilde{\Gamma}^i_{jk}$ is the Christoffel symbol for the metric
$H_{ij}$. It means that a configuration of a stationary string in a static
spacetime
coincides with a geodesic for the 3-dimensional "unphysical" space with
the line-element:
\begin{equation}
d\tilde{l}^2=H_{ij}dx^i dx^j=Fdl^2,
\end{equation}
where $dl$ is the physical distance in the static spacetime defined by Eq.
(3.1)
and $\sigma$ is an affine parameter (proper length):
\begin{equation}
H_{ij}x'^i x'^j=1.
\end{equation}
Here and later on
a tilde indicates that an object is defined with respect to the 3-dimensional
"unphysical"
space (3.8). We use now this reduction to simplify the equations for the
strings perturbations. To be more concrete we express the effective
action (2.18) for the perturbations in terms of quantities defined in
the "unphysical" space.

Consider the 2 normalvectors $n^\mu_R$ introduced in (2.7). Eqs. (3.4) and
(2.7) imply that
$n^\mu_R$ are of the form:
\begin{equation}
n^\mu_R=(0,n^i_R),
\end{equation}
so that we can exclude from our consideration the zero-components and consider
$(x'^i,n^i_2,n^i_3)$
as an orthogonal system in the 3-dimensional "unphysical" space. The
normalvectors are however
not normalized with respect to $H_{ij}$ and it is therefore convenient to
rewrite the decompositions of the
perturbations $\delta x^\mu$ in the following way:
\begin{equation}
\delta x^i=n^i_R\delta x^R=(\frac{n^i_R}{\sqrt{F}})(\sqrt{F}\delta x^R)\equiv
\tilde{n}^i_R\tilde{\delta x}^R,\hspace*{5mm}\delta x^0=0.
\end{equation}
In this case $(x'^i,\tilde{n}^i_2,\tilde{n}^i_3)$ is an orthonormal system
in the 3-dimensional space with metric $H_{ij}$:
\begin{equation}
H_{ij}\tilde{n}^i_R\tilde{n}^j_S=\delta_{RS},\hspace*{5mm}H_{ij}x'^i
\tilde{n}^j_R=0,
\end{equation}
remembering that $x'^i$ is already normalized because of Eq. (3.9).

The Christoffel symbols $\Gamma^0_{i0}$ and $\Gamma^i_{00}$
are given by Eq. (3.2). The conformal relation between the
spacelike components of $g_{\mu\nu}$ and $H_{ij}$ leads to:
\begin{equation}\Gamma^i_{jk}=\tilde{\Gamma}^i_{jk}-\frac{1}{2F}(F_{,j}
\delta^i_k+F_{,k}\delta^i_j-F_{,l}H^{li}H_{jk}).
\end{equation}
{}From this equation and Eqs. (2.10)-(2.11) we can now easily obtain the second
fundamental
form $\Omega_{R,AB}$ and the normal fundamental form $\mu_{RS,A}$ in terms
of quantities defined in the unphysical space:
\begin{equation}
\Omega_{R,01}=\Omega_{R,10}=0,
\end{equation}
\begin{equation}
\Omega_{R,00}=F^2\Omega_{R,11}=\frac{\sqrt{F}}{2}\tilde{n}^i_R F_{,i}
\end{equation}
as well as:
\begin{equation}
\mu_{RS,0}=0,
\end{equation}
\begin{equation}
\mu_{RS,1}=H_{ij}\tilde{n}^i_R x'^k\tilde{\nabla}_k\tilde{n}^j_S
\end{equation}

Finally we need a relation between the Riemann tensors. Generally it turns out
to be very complicated but fortunately we only need the special projection
appearing in equations (2.14) and (2.18)-(2.19), which after some tedious but
straightforward algebra leads to:
\begin{eqnarray}
\delta x^R h^{AB}x^\mu_{,A} x^\nu_{,B}R_{\mu\rho\sigma\nu}n^\rho_R n^\sigma_S
\delta x^S=\tilde{\delta x}^R (x'^i x'^j
\tilde{R}_{iklj}\tilde{n}^k_R\tilde{n}^l_S\hspace*{-2mm}&+&
\hspace*{-2mm}\frac{F_{,k}F_{,l}}{2F^2}
\tilde{n}^k_R
\tilde{n}^l_S)\tilde{\delta x}^S\nonumber \\
+\tilde{\delta x}_R x'^i x'^j (\frac{
\tilde{\Gamma}^k_{ij}}{2F}F_{,k}-
\frac{F_{,ij}}{2F}\hspace*{-2mm}&+&\hspace*{-2mm}\frac{F_{,i}F_{,j}}{4F^2})
\tilde{\delta x}^R.
\end{eqnarray}
Collecting everything we get the effective action Eq. (2.18) in the form:
\begin{eqnarray}
{\cal S}_{eff}=\int d\tau d\sigma \tilde{\delta x}^R(\hspace*{-2mm}
&\delta_{RS}&\hspace*{-2mm}(\partial^2_\sigma-\frac{\partial^2_\tau}{F^2})-x'^i
x'^j
\tilde{R}_{iklj}\tilde{n}^k_R\tilde{n}^l_S\nonumber \\
\hspace*{-2mm}&+&\hspace*{-2mm}2\mu_{RS,1}\partial_\sigma-\mu_{RT,1}
\mu_S\hspace*{1mm}^T\hspace*{1mm}_1)\tilde{\delta x}^S,
\end{eqnarray}
where we also used (from Eq. (2.4)):
\begin{equation}
\Box\equiv -\frac{\partial^2_\tau}{F}+F\partial^2_\sigma+F'\partial_\sigma.
\end{equation}
This action allows further simplification by using the gauge-freedom
related to the choice of the normalvectors discussed at the end of section 2.
To fix this freedom we choose the basis $(x'^i,\tilde{n}^i_2,\tilde{n}^i_3)$
obeying the normalization conditions (3.12) at a given point, and define the
basis along the geodesic by means of parallel transport. For such
a choice $\mu_{RS,1}=0$ and Eq. (3.19) reduces to:
\begin{equation}
{\cal S}_{eff}=\int d\tau d\sigma \tilde{\delta
x}^R(\delta_{RS}(\partial^2_\sigma-
\frac{\partial^2_\tau}{F^2})-x'^i x'^j\tilde{R}_{iklj}\tilde{n}^k_R
\tilde{n}^l_S)\tilde{\delta x}^S,
\end{equation}
with corresponding equations of motion for $\tilde{\delta x_R}$:
\begin{equation}
(\partial^2_\sigma-\frac{\partial^2_\tau}{F^2})\tilde{\delta x}_R=V_{RS}
\tilde{\delta x}^S,
\end{equation}
where the matrix-potential $V_{RS}$ is given by:
\begin{equation}
V_{RS}=x'^i x'^j\tilde{R}_{iklj}\tilde{n}^k_R\tilde{n}^l_S.
\end{equation}
There is a simple relation between Eqs. (3.22)-(3.23)
and the so-called geodesic deviation equation. The geodesic deviation equation
describing
the rate of spread of 2 neighbouring geodesics is in our notation
given by [15]:
\begin{equation}
x'^j\tilde{\nabla}_j(x'^k\tilde{\nabla}_k\delta x^i)=\tilde{R}^i_{jkl}x'^j
x'^k\delta x^l.
\end{equation}
Using Eq. (3.11) and multiplying both sides by $H_{ij}\tilde{n}^j_S$ gives:
\begin{equation}
\frac{d^2}{d\sigma^2}\tilde{\delta x}_R+H_{ij}\tilde{n}^i_S
[x'^k\tilde{\nabla}_k(
x'^l\tilde{\nabla}_l\tilde{n}^j_R)+2(x'^k\tilde{\nabla}_k\tilde{n}^j_R)
\frac{d}{d\sigma}]
\tilde{\delta x}^S=x'^i x'^j\tilde{R}_{iklj}\tilde{n}^k_R
\tilde{n}^l_S\tilde{\delta x}^S.
\end{equation}
With our special choice of the normalvectors the terms in the square bracket
vanish and the result is equivalent to Eq. (3.22) except for the
$\partial_\tau$-term. For time-independent perturbations our result
therefore leads to the geodesic deviation equation, as it of course should, but
due to the
possibility of having time-dependent perturbations we get an extra term. This
extra term can be written in a more convenient form if we re-introduce the
variations
$\delta x^R=\tilde{\delta x}^R/\sqrt{F}$ (c.f. Eq. (3.11)) and define what we
will
call the conformal string-parameter $\sigma_c$:
\begin{equation}
d\sigma_c=\frac{d\sigma}{F}.
\end{equation}
In this case Eq. (3.22) becomes:
\begin{equation}
(\partial^2_{\sigma_c}-\partial^2_\tau)\delta x_R=U_{RS}\delta x^S,
\end{equation}
where the potential $U_{RS}$ is given by:
\begin{equation}
U_{RS}=V\delta_{RS}+F^2 V_{RS}.
\end{equation}
Here $V_{RS}$ is the potential (3.23) while $V$ is given by:
\begin{equation}
V=\frac{3}{4F^2}\left( \frac{dF}{d\sigma_c}\right) ^2-\frac{1}{2F}\frac
{d^2 F}{d\sigma^2_c}=\frac{1}{4}(F'^2-2FF'').
\end{equation}
Equations (3.27)-(3.29) finally represent the desired results of this section.
Since we consider time-dependent perturbations propagating along a
stationary string in a static
background, we can split the $\tau$ and $\sigma_c$-dependence of $\delta x^R$
so that Eq. (3.27) leads to a Schr\"{o}dinger-like matrix equation. The
potential $U_{RS}$
consists of 2 terms: The first (diagonal) term $V\delta_{RS}$ is defined by
the red-shift factor $F$ and it is connected with the time-delay effect in a
static gravitational field. The second (generally non-diagonal) term $V_{RS}$
is connected with the curvature of the 3-dimensional "unphysical" space.

We conclude this section with the following remark. In the general case the
presence of the spacetimes curvature may result in the mixture of polarizations
of the perturbations during their propagation along the string. This mixture
may be absent if the spacetime under consideration possesses symmetries. Let
us suppose that the metric $H_{ij}$ is invariant under the discrete symmetry
transformation $(x^1,x^2,x^3)\longrightarrow (x^1,x^2,-x^3)$. It is evident
that for a static string lying in the plane $x^3=0$ the perturbations in the
$x^3$-direction (perpendicular to the string) and in the strings plane cannot
be mixed. For the corresponding choice of $\tilde{n}_2^i=n_\perp^i$ and
$\tilde{n}_3^i=n_\parallel^i$ the potential $U_{RS}$ in Eq. (3.27) becomes
diagonal.

In the next section we will consider equation (3.27) in a few special cases.
\vskip 6pt
\section{Special cases}
\subsection{Quasi-Newtonian gravitational field}
\setcounter{equation}{0}
\vskip 6pt
As a first example we consider the perturbations propagating along a string
in a static gravitational field
described in the quasi-Newtonian approximation. In this
approximation the line-element is [12]:
\begin{equation}
ds^2=-(1+2\phi)dt^2+(1-2\phi)(dx^2+dy^2+dz^2),
\end{equation}
where $\phi=\phi (x,y,z)$ is a gravitational potential.

In the notation of Eq. (3.1) we find:
\begin{equation}
F=1+2\phi,\hspace*{5mm}H_{ij}=(1-2\phi)(1+2\phi)\delta_{ij}\approx\delta_{ij},
\end{equation}
i.e. the 3-dimensional "unphysical" space is flat and the stationary string is
represented simply by a straight line, according to Eq. (3.7). As the result
the term $V_{RS}$ in the effective potential $U_{RS}$ (3.27) vanishes and the
equation
for the propagation of perturbations takes the form:
\begin{equation}
(\partial^2_{\sigma_c}-\partial^2_\tau)\delta x_R=V\delta x_R,
\end{equation}
where $V=-d^2\phi/{d\sigma_c^2}$ to first order in $\phi (x^i)$.
After Fourier-expanding $\delta x_R$:
\begin{equation}
\delta x_R (\tau,\omega_c)=\int e^{-i\omega\tau} D_\omega^R (\sigma_c) d\omega,
\end{equation}
equation (4.3) reduces to:
\begin{equation}
\left(\frac{d^2}{d\sigma_c^2}+\omega^2-V\right)
D_\omega^R=0.
\end{equation}
In the adiabatic approximation we formally obtain the solution to this equation
in the form:
\begin{equation}
D_\omega^R(\sigma_c)=e^{\pm i\int^{\sigma_c}\sqrt{\omega^2-V}d\sigma'_c}.
\end{equation}
\vskip 6pt
\subsection{Rindler space}
\vskip 6pt
In this subsection we consider a stationary string in Rindler space,
corresponding
to a static homogeneous gravitational background. This case, and the equivalent
problem of a uniformly accelerating string in Minkowski space, has earlier been
investigated from a different point of view in Ref.4.  The line-element of
Rindler space may be written [16]:
\begin{equation}
ds^2=-a^2x^2 dt^2 +dx^2+dy^2+dz^2,
\end{equation}
where $a$ is a positive constant.

In the notation of section 3 we can then identify:
\begin{equation}
F=a^2 x^2,\hspace*{5mm}H_{ij}=a^2 x^2 \delta_{ij},
\end{equation}
i.e. the "unphysical" space is conformally flat. The corresponding Christoffel
symbols and Riemann tensor components are found to be:
\begin{equation}
\tilde{\Gamma}^x_{xx}=-\tilde{\Gamma}^x_{yy}=-\tilde{\Gamma}^x_{zz}=
\tilde{\Gamma}^y_{xy}=\tilde{\Gamma}^z_{xz}=\frac{1}{x},
\end{equation}
as well as:
\begin{equation}
\tilde{R}_{xyxy}=\tilde{R}_{xzxz}=-\tilde{R}_{yzyz}=a^2.
\end{equation}
Next we have to solve the geodesic equation (3.7) in order to determine the
stationary string configuration. In order to exclude possible misunderstanding
we emphasize that a free string cannot be at rest in the Rindler spacetime. In
order to be able to study a stationary configuration one needs to assume that
there are additional (non-gravitational) forces acting either on the string
or on its endpoints. We do not consider the delicacies concerning this point
here but refer to Ref. [4].
Due to the symmetry of the problem we may consistently
take $z=0$. After one integration the other 2 equations of (3.7) then lead to:
\begin{equation}
y'=\frac{b}{x^2},\hspace*{5mm}x'^2+\frac{b^2}{x^4}=\frac{1}{a^2 x^2},
\end{equation}
where $b$ is an integration constant. These equations can be easily solved
by introducing the conformal string parameter (3.26). The solution reads:
\begin{equation}
x(\sigma_c)=ba\cosh (a\sigma_c),\hspace*{5mm}y(\sigma_c)=ba^2\sigma_c.
\end{equation}

We choose the 2 normalvectors $n_\perp^i$ and $n_\parallel^i$ in the following
way:
\begin{equation}
n_\perp^i=\frac{1}{a x}(0,0,1),\hspace*{5mm}n_\parallel^i=(-y',x',0).
\end{equation}
(The prime as earlier denotes the derivative with respect to $\sigma$.)
It can be shown that $n_\perp^i$ and $n_\parallel^i$ are covariantly constant
in
the 3-dimensional "unphysical" space.
Note that $n_\perp^i$ is pointing in the $z$-direction perpendicular to the
plane of the string while $n_\parallel^i$ is lying in the plane of the
string, c.f. the discussion at the end of section 3.

To determine the time-dependent perturbations of the string in the 2
directions we have to calculate the components of the potential $U_{RS}$.
Using Eqs. (4.8) and (4.10)-(4.13) one finds:
\begin{equation}
V_{\perp\perp}=\frac{2b^2 a^2-x^2}{a^2 x^6},\hspace*{5mm}
V_{\parallel\hspace*{.5mm}\parallel}=\frac{-1}{a^2 x^4},
\hspace*{5mm}V_{\perp\parallel}=V_{\parallel\perp}=0,\hspace*{5mm}
V=\frac{a^2}{x^2}(x^2-2a^2 b^2).
\end{equation}
The equations of motion for the perturbations (3.27) become:
\begin{equation}
(\partial^2_{\sigma_c}-\partial^2_\tau)\delta x_\perp=0,
\end{equation}
\begin{equation}
(\partial^2_{\sigma_c}-\partial^2_\tau)\delta x_\parallel=
-\frac{2a^4 b^2}{x^2}\delta x_\parallel=-\frac{2a^2}{\cosh^2 (a\sigma_c)}
\delta x_\parallel.
\end{equation}

The perturbations connected with the oscillations of the string
in the direction
perpendicular to its plane are
described by simple plane waves in the $(\tau,\sigma_c)$-coordinates,
while the
perturbations in the plane of the string (4.16) are somewhat more
complicated. After
Fourier expanding $\delta x_\parallel$:
\begin{equation}
\delta x_\parallel (\tau, \sigma_c)= \int e^{-i\omega\tau} D_\omega
(\sigma_c)d\omega,
\end{equation}
equation (4.16) leads to:
\begin{equation}
\frac{d^2}{d\sigma^2_c}D_\omega+\left( \omega^2+\frac{2 a^2}
{\cosh^2 (a\sigma_c)}\right)D_\omega=0,
\end{equation}
which is a well-known equation in quantum mechanics; the so-called
P\"{o}schl-Teller equation [17]. The complete solution may be written in terms
of
hypergeometric functions but we shall not go into any details here. It suffices
to say that if we consider a scattering-process, i.e. we consider solutions of
the asymptotic form:
\begin{equation}
D_\omega (\sigma_c) = \left\{ \begin{array}{cl}
e^{i\omega\sigma_c}+R_\omega e^{-i\omega\sigma_c} & \mbox{for
$\sigma_c\rightarrow
-\infty$} \\
T_\omega e^{i\omega\sigma_c} & \mbox{for $\sigma_c\rightarrow\infty$}
\end{array}
\right.,
\end{equation}
then it can be shown that the reflection-coefficient $R_\omega$ vanishes [17].
The only
effect of the potential is then the appearence of a phase factor in the
transmitted
wave $T_\omega\equiv e^{2i\phi_e}$ [17]:
\begin{equation}
\phi_e (\omega)=\arg\left( \frac{\Gamma
(\frac{i\omega}{a})e^{-i\frac{\omega}{a}\log 2}}
{\Gamma (1+\frac{i\omega}{2a})\Gamma (\frac{i\omega}{2a})}\right).
\end{equation}
\vskip 6pt
\subsection{de Sitter space}
\vskip 6pt
As a third example we consider a slightly more complicated case, namely the
stationary string in the de Sitter space. We use static coordinates in which
the de Sitter metric takes the form:
\begin{equation}
ds^2=-f(r)dt^2+\frac{dr^2}{f(r)}+r^2(d\theta^2+\sin^2\theta d\phi^2).
\end{equation}
Here $f(r)=1-H^2 r^2$, and $H$ is the constant Hubble-parameter. In this
case we find:
\begin{equation}
F=f(r),\hspace*{5mm}H_{ij}=diag(1,r^2 f(r),r^2 f(r) \sin^2\theta),
\end{equation}
with the following expressions for non-vanishing components of the Christoffel
symbols $\tilde{\Gamma}^i_{jk}$ and Riemann tensor $\tilde{R}_{ijkl}$:
\begin{eqnarray}
\tilde{\Gamma}^r_{\phi\phi}\hspace*{-2mm}&=&\hspace*{-2mm}\sin^2\theta
\tilde{\Gamma}^r_{\theta\theta}=-r(1-2H^2 r^2)\sin^2\theta,\hspace*{5mm}
\tilde{\Gamma}^\phi_{\theta\phi}=\cot\theta,\nonumber\\
\tilde{\Gamma}^\phi_{r\phi}\hspace*{-2mm}&=&\hspace*{-2mm}\tilde
{\Gamma}^\theta_{r\theta}=\frac{1-2H^2 r^2}{rf(r)},\hspace*{5mm}\tilde
{\Gamma}^\theta_{\phi\phi}=-\cos\theta\sin\theta
\end{eqnarray}
and:
\begin{eqnarray}
\tilde{R}_{r\phi r\phi}\hspace*{-2mm}&=&\hspace*{-2mm}\sin^2\theta\tilde
{R}_{r\theta r\theta}=\frac{H^2 r^2 (3-2H^2 r^2)}{f(r)}\sin^2\theta,\nonumber\\
\tilde{R}_{\theta\phi\theta\phi}\hspace*{-2mm}&=&\hspace*{-2mm}
H^2 r^4 (3-4H^2 r^2).
\end{eqnarray}
Consider now the geodesic equation (3.7). Without loss of generality we assume
that the string lies in the equatorial
plane $(\theta=\pi/2)$ and find after one integration:
\begin{equation}
\phi'=\frac{bH^2}{r^2 f(r)},
\end{equation}
\begin{equation}
r'^2=1-\frac{b^2 H^4}{r^2 f(r)},
\end{equation}
with $b$ an integration constant. The normalvectors $n_\perp^i$ and
$n_\parallel^i$
are conveniently
chosen in the form:
\begin{equation}
n_\perp^i=\frac{1}{r\sqrt{f(r)}}(0,1,0),\hspace*{5mm}n_\parallel^i=
\frac{1}{r\sqrt{f(r)}}(-bH^2,0,r').
\end{equation}
As in the case of Rindler space (Section 4.2) for this choice $n_\parallel^i$
lies in the plane of the string while $n_\perp^i$ is perpendicular to this
plane.
Note also that $n_\parallel^i$ and $n_\perp^i$
are covariantly constant in the "unphysical" space. The components
of the potential $U_{RS}$ read:
\begin{eqnarray}
V_{\perp\perp}\hspace*{-2mm}&=&\hspace*{-2mm}\frac{H^2}{f^3 (r)}
(-3+2H^6 b^2+5H^2 r^2-
2H^4 r^4),\hspace*{5mm}V_{\perp\parallel}=V_{\parallel\perp}=0,\nonumber\\
V_{\parallel\hspace*{.5mm}\parallel}\hspace*{-2mm}&=&\hspace*{-2mm}-\frac{H^2}{f^2 (r)}(3-2H^2 r^2),
\hspace*{5mm}V=H^2 (1-\frac{2b^2 H^6}{f(r)}).
\end{eqnarray}
The equations of motion for the time-dependent perturbations are of the
form (3.27):
\begin{equation}
(\partial^2_{\sigma_c}-\partial^2_\tau)\delta x_\perp=-2H^2 f(r)\delta x_\perp,
\end{equation}
\begin{equation}
(\partial^2_{\sigma_c}-\partial^2_\tau)\delta x_\parallel=-2H^2 (f(r)+\frac{b^2
H^6}
{f(r)})\delta x_\parallel,
\end{equation}
where $f(r)=1-H^2 r^2 (\sigma_c)$, and (from Eqs. ( 3.26) and (4.26)):
\begin{equation}
r^2 (\sigma_c)=\frac{1}{H^4}\wp (\sigma_c+z_o)+\frac{2}{3 H^2}.
\end{equation}
Here $\wp (z)$ is the Weierstrass elliptic $\wp$-function [18] with invariants:
\begin{equation}
g_2=4 H^4 (\frac{1}{3}-b^2 H^6),\hspace*{5mm}g_3=\frac{4}{3}H^6 (-\frac{2}{9}
+b^2 H^6).
\end{equation}
and $z_o$ is an integration constant which specifies the solution.

Let us now consider 2 special configurations (4.31) where we can write down
explicit analytic solutions to Eqs. (4.29)-(4.30).

For $2bH^3=1$ (4.31) reduces to:
\begin{equation}
r^2 (\sigma_c)=\frac{1}{2H^2},
\end{equation}
which is just a circular string with constant radius. In this case the
perturbations
are simply determined by:
\begin{equation}
(\partial^2_{\sigma_c}-\partial^2_\tau)\delta x_\perp+H^2\delta x_\perp=0,
\end{equation}
\begin{equation}
(\partial^2_{\sigma_c}-\partial^2_\tau)\delta x_\parallel+
2H^2\delta x_\parallel=0.
\end{equation}
{}From Eqs. (3.26), (4.25) and (4.33) follows that:
\begin{equation}
\phi=H\sigma_c,
\end{equation}
so that $\sigma_c$ is periodic with period $2\pi/H$. This periodicity is to be
considered as boundary conditions for the solutions to Eqs. (4.34)-(4.35),
which are
then represented by plane waves in the form ($n\in Z$):
\begin{equation}
\delta x_\perp\sim e^{-iH(\tau\sqrt{n^2-1}\pm n\sigma_c)},
\end{equation}
\begin{equation}
\delta x_\parallel\sim e^{-iH(\tau\sqrt{n^2-2}\pm n\sigma_c)},
\end{equation}
so that the perturbations in the 2 normal directions propagate with
different frequencies (for fixed wavenumber). Note that there are unstable
modes
also, corresponding to imaginary frequencies. This is easy to understand
physically since the stationary circular string needs an exact balancing of the
string tension and the expansion of the universe. Considering for instance
the perturbation in the plane of the string (4.38), we find that the
$(n=1)$-mode
corresponds to a wavelength that is larger than the circumference of
the string (using Eq. (4.33)):
\begin{equation}
\lambda=\frac{2\pi}{Hn}\mid_{n=1}=\frac{2\pi}{H}=\sqrt{2}(2\pi r).
\end{equation}
For such perturbations the string will either collapse to a point or expand
towards the horizon.

As another simple configuration in de Sitter space we take a "straight" string
which is obtained by the choice $b=0$. In this case the Weierstrass function
in Eq. (4.31) reduces to a hyperbolic function:
\begin{equation}
r^2 (\sigma_c)=\frac{1}{H^2}\tanh^2 (H\sigma_c).
\end{equation}
The equations for the 2 different polarizations of the perturbations are
now identical and read:
\begin{equation}
(\partial^2_{\sigma_c}-\partial^2_\tau)\delta x_R+\frac{2H^2}
{\cosh^2 (H\sigma_c)}\delta x_R=0.
\end{equation}
After Fourier expansion:
\begin{equation}
\delta x_R (\tau,\sigma_c)=\int e^{-i\omega\tau} D^R_\omega (\sigma_c)d\omega,
\end{equation}
we re-discover the P\"{o}schl-Teller equation (4.18) with $a$ replaced by $H$:
\begin{equation}
\frac{d^2}{d\sigma^2_c} D^R_\omega+\left( \omega^2+\frac{2H^2}{\cosh^2
(H\sigma_c)}
\right) D^R_\omega=0,
\end{equation}
and with similar conclusions as after (4.18).
\vskip 6pt
\subsection{Black holes}
\vskip 6pt
As a final example we now consider a stationary string in the background of
a black hole. The stationary configurations of the string in the black hole
metrics were described in Ref. [11]. For a stationary string lying in the
equatorial plane ($\theta=\pi/2$) of a Schwarzschild black hole with the
metric:
\begin{equation}
ds^2=-(1-\frac{2m}{r})dt^2+(1-\frac{2m}{r})^{-1} dr^2+r^2 (d\theta^2+
\sin^2\theta d\phi^2),
\end{equation}
where $m$ is the mass of the black hole, $r=r(\sigma_c)$ is defined by the
equation:
\begin{equation}
\left(\frac{dr}{d\sigma_c}\right)^2=\frac{\Delta(\Delta-b^2)}{r^4}.
\end{equation}
Here $\Delta=r^2-2mr$ and $b$ is an integration constant.

The calculation of the various parts of the potential $U_{RS}$
proceeds in the same way as in subsections 4.1-4.3. The details are however
a little more involved due to the complexity of the Christoffel symbols,
Riemann tensors etc, so we just give here the result. The two polarizations
of perturbations (parallel and perpendicular to the strings plane) propagate
independently. The corresponding equations read:
\begin{equation}
(\partial^2_{\sigma_c}-\partial^2_\tau)\delta x_\perp=\frac{m}{r^5}
(2\Delta-3b^2)\delta x_\perp,
\end{equation}
\begin{equation}
(\partial^2_{\sigma_c}-\partial^2_\tau)\delta x_\parallel=
[\frac{m}{r^5}(2\Delta-3b^2)-\frac{2m^2 b^2}{\Delta r^4}]\delta x_\parallel.
\end{equation}
The solution of these equations is supposed to be analyzed somewhere else.
\vskip 6pt
The above considerations can be easily generalized to the case of a charged
black hole. We shall not present here the results but restrict ourselves by the
following interesting observation. The equations for the string perturbations
as well as the equation for the equilibrium configuration of a string are
greatly simplified for an extreme
Reissner-Nordstr\"{o}m
black hole. The line-element for this case is:
\begin{equation}
ds^2=-\frac{(r-m)^2}{r^2}dt^2+\frac{r^2}{(r-m)^2}dr^2+r^2 (d\theta^2+
\sin^2\theta d\phi^2),
\end{equation}
where $m$ is the mass of the black hole (the charge $Q=m$). It follows that:
\begin{equation}
F=\frac{(r-m)^2}{r^2},\hspace*{5mm}H_{ij}=diag(1,(r-m)^2,(r-m)^2\sin^2\theta).
\end{equation}
$H_{ij}$ is therefore the metric of ordinary spherical coordinates, with
$(r-m)$
being the radial coordinate, i.e. the 3-dimensional "unphysical" space is flat
and the
stationary string is represented by a straight line [11]. This is seen
explicitly
by solving equation (3.7), which in the equatorial plane $(\theta=\pi/2)$ leads
to:
\begin{equation}
\phi'=\frac{b}{(r-m)^2},\hspace*{5mm}r'^2=1-\frac{b^2}{(r-m)^2},
\end{equation}i.e.:
\begin{equation}
\phi (r-m)=\pm\arcsin\left( \frac{\sqrt{(r-m)^2-b^2}}{r-m}\right),
\end{equation}
where $b$ is an integration constant. The potential $U_{RS}$ which enters
in the equations of motion for the
perturbations (3.27) contains only the term from the time-dilation part:
\begin{equation}
V=\frac{m}{r^6}(2(r-m)^3-b^2 (3r-2m)).
\end{equation}
It follows that the perturbations are identical for both polarizations
(perpendicular to the plane of the string and in the plane of the string). The
solutions of Eq. (3.27) with the potential (4.52) are to be discussed
elsewhere.
\vspace*{2cm}\\
{\bf Acknowledgements}
\vskip 6pt
This work was supported in part by the Danish Natural Science Research
Council Grant No. 11-9524-1SE and by Deutsche Forschungsgemeinschaft (V.P.F).
\newpage
\begin{centerline}
{\bf References}
\end{centerline}
\vskip 6pt
\begin{enumerate}
\item V.E. Zakharov and A.V. Michailov, JETP 47 (1979) 1017;
      H. Eichenherr, {\it in}: Integrable quantum field theories, Lecture
      Notes in Physics, vol. 151, Tv\"{a}rminne Proc., ed. J. Hietarinta and
      C. Montonen (Springer-Verlag, Berlin, 1982)
\item H.J. De Vega, M. Ramon-Medrano and N. S\'{a}nchez, Nucl. Phys. B374
(1992) 405
\item H.J. De Vega and N. S\'{a}nchez, Nucl. Phys. B317 (1989) 706, 731
\item V.P. Frolov and N. S\'{a}nchez, Nucl. Phys. B349 (1991) 815
\item J. Garriga and A. Vilenkin, Phys. Rev. D44 (1991) 1007
\item A. Vilenkin, Phys. Rep. 121 (1985) 264
\item A. Vilenkin, Phys. Rev. D41 (1990) 3038;
      J. Hong, J. Kim and P. Sikivie, Phys. Rev. Lett. 69 (1992) 2611;
      B. Carter, Phys. Rev. D41 (1990) 3869
\item B. Carter, V.P. Frolov and O. Heinrich, Class. Quant. Grav. 8 (1991) 135
\item H.J. De Vega and N. S\'{a}nchez, Phys. Lett. B197 (1987) 320
\item A.M. Polyakov, Phys. Lett. B103 (1981) 207
\item V.P. Frolov, V.D. Skarzhinsky, A.I. Zelnikov and O. Heinrich,
      Phys. Lett. B224 (1989) 255
\item C.W. Misner, K.S. Thorne and J.A. Wheeler, Gravitation (Freeman,
      San Francisco, CA, 1973)
\item L.P. Eisenhart, Riemannian Geometry (Princeton University Press, fifth
      printing, 1964) Chapter IV
\item Y. Nambu, Lectures at the Copenhagen Summer Symposium (1970);
      T. Goto, Prog. Theor. Phys. 46 (1971) 1560
\item L.D. Landau and E.M. Lifshitz, The Classical Theory of Fields (Pergamon
      Press, fourth english edition, 1975) p. 262
\item W. Rindler, Essential Relativity (Springer-Verlag, New York, second
      edition, 1977) p. 156
\item S. Fl\"{u}gge, Practical Quantum Mechanics I (Springer-Verlag, New York,
      1971) p. 94-99
\item M. Abramowitz and I.A. Stegun, Handbook of Mathematical Functions (Dover
      Publications Inc, New York, ninth printing) Chapter 18

\end{enumerate}
\end{document}